\documentclass{elsart}

\usepackage{amssymb}
\usepackage{graphicx}
\usepackage{rotating}
\usepackage{tikz}
\begin{document}

\begin{frontmatter}
\title{Analysis of proton and 
neutron pair breakings: High-spin structures of $^{124-127}$Te isotopes}
\author{}
\author{Vikas Kumar$^a$,}
\author{P. C.~Srivastava$^a$,}
\author{M. J. Ermamatov$^{b,c}$, }
\author{ and Irving O. Morales$^d$}
\address{$^a$Department of Physics, Indian Institute of Technology, Roorkee 247 667, 
India}
\address{$^b$ Instituto de Fis\'ika, Universidade Fideral Fluminense, 24210-340,
Niter{\'o}i, Rio de Janeiro, Brazil }
\address{$^c$Institute of Nuclear Physics, Ulughbek, Tashkent 100214, Uzbekistan} 
\address{$^d$ Instituto de Ciencias Nucleares, Universidad Nacional Aut\'onoma de M\'exico, 04510 M\'exico, D.F., Mexico}

\date{\today}

\begin{abstract}  
In the present work recently available experimental data  for high-spin states 
of four nuclei, $^{124}_{\ 52}$Te, $^{125}_{\ 52}$Te, $^{126}_{\ 52}$Te, and $^{127}_{\ 52}$Te have been interpreted using
state-of-the-art shell model calculations. The calculations have been performed in the $50-82$ valence shell composed of  
$1g_{7/2}$, $2d_{5/2}$, $1h_{11/2}$, $3s_{1/2}$, and $2d_{3/2}$ orbitals. We have compared our results with the available experimental data for
excitation energies and transition probabilities, including high-spin states. The results are in reasonable agreement with the available experimental data.
The wave functions, particularly, the specific proton and neutron configurations which are involved to generate the angular momentum along the yrast lines are discussed. 
 We have also estimated overall contribution of three-body forces in the energy level shifting. Finally, results with modified effective interaction are also reported.

\end{abstract}


\end{frontmatter}

\section{Introduction}

Neutron rich nuclei in the vicinity of Sn ($Z=50$) are important in many ways, viz. abundance of isomeric states,
a possible candidate of neutrinoless double beta decay ($^{124}$Sn), astrophysical interest etc \cite{nature,prc1,prc2,prc3,prc4,wood,prl1,prl2,prl3}. 
The high-seniority states in the case of Sn isotopes with triple pair breaking (seniority $v$=6), have been reported in the
literature \cite{talmi,astierconf}. This region is also important to test
nuclear models for the correct prediction of nuclear spectroscopic properties. Beyond Sn  many
experimental works have recently been done to investigate the nuclear structure properties of Te and Xe isotopes. 

Recently Astier {\it et al.} \cite{teepja}  populated $^{124-131}$Te isotopes as fission fragments in two fusion-fission reactions $^{18}$O + $^{208}$Pb and $^{12}$C + $^{238}$U induced by heavy ions,  using Euroball array. In this experiment high-spin level schemes extended up to 6 MeV (for even-Te)
and 5 MeV (for odd-Te). The yrast excitations in $A=126-131$ Te isotopes from deep inelastic $^{130}$Te+$^{64}$Ni reactions were reported in ref. \cite{zhang}.
In this work, the information especially on yrast excitations in the odd-A $^{127,129,131}$Te isotopes is discussed. Both single-particle
and collective aspects of the level spectra are analyzed there.

High-spin states of $^{136}$Xe, $^{137}$Cs, $^{138}$Ba,
$^{139}$La and $^{140}$Ce are populated by  Astier {\it et al.} \cite{PhysRevC.85.064316} for $N=82$ isotones. In the frame work of shell model we have interpreted these experimental data successfully for these nuclei in ref.~\cite{JPGN82}.
In this mass region previously one of us have analyzed experimentally observed slow $E3$ transition in $^{136}$Cs ~\cite{PhysRevC.84.014329}
which was populated 
at ISOLDE facility at CERN and also for the high-spin states of $^{136}$Cs ~\cite{PhysRevC.87.054316} which were populated by XTU Tandem accelerator
at Legnaro and Vivitron accelerator of IRes, Strasbourg. The
  $B(E2)$ transition trends have recently been studied by I. O. Morales
{\it et al}~\cite{Morales2011606} in tin isotopes using generalized seniority approach.

As we approach towards the $N=82$ shell, the mechanism of the formation of high-spin states starts to differ in Sn ($Z=50$) to Te ($Z=52$). In the case of Sn isotopes \cite{broda,Pietri},
this is due to breaking of neutrons pairs, while for Te isotopes,
various configurations are expected from the breaking of proton/neutron pairs near the subshells that includes neutrons
in $h_{11/2}$, $d_{3/2}$ orbitals and protons in $g_{7/2}$ and $d_{5/2}$ orbitals. Thus, there is a competition of protons and neutrons pair breaking for high-spin states in Te isotopes.


The aim of the present work is to discuss shell model results of newly populated high-spin states of $^{124-127}$Te isotopes \cite{teepja}.
This work will add more information to the work by Astier {\it et al}  \cite{teepja} on Te isotopes, where shell model results only for $^{128-131}$Te isotopes were reported. 

This work is organized as follows:  comprehensive
comparison of shell-model results and experimental data are given in Section 2. In Section 3, 
transition probabilities are compared with the available experimental data.   
In section 4, we have estimated contributions from three-body forces. Results with modified interaction are shown in section 5.
Finally, concluding remarks are drawn in Section 6.
 
\section{Shell model results and discussions}

The shell-model calculations for the Te isotopes have been
performed in the 50-82 valence shell composed of the orbits $1g_{7/2}$, $2d_{5/2}$,
$1h_{11/2}$, $3s_{1/2}$, and $2d_{3/2}$.  We have performed calculations with
SN100PN interaction due to Brown {\it et al}~\cite{Mac,PhysRevC.71.044317}. This interaction has four parts: 
neutron-neutron, neutron-proton, proton-proton and Coulomb repulsion between the protons.
The single-particle energies for the neutrons
are -10.6089, -10.289, -8.717, -8.694, and -8.816 MeV
for the $1g_{7/2}$, $2d_{5/2}$, $2d_{3/2}$, $3s_{1/2}$, and $1h_{11/2}$ orbitals, respectively,
and those for the protons are 0.807, 1.562, 3.316, 3.224,
and 3.605 MeV. The results shown in this work were obtained with the code NuShellX \cite{MSU-NSCL}.
In the present work we have employed truncation for the neutron orbitals.
For $^{124,126}$Te, we filled completely $\nu g_{7/2}$ orbital and put minimum 4 particles in $\nu d_{5/2}$ orbital. 
In the case of $^{125,127}$Te, we slightly relaxed truncation by filling completely $\nu g_{7/2}$ orbital and putting only minimum 2 
particles in $\nu d_{5/2}$ orbital. We have calculated the electromagnetic properties of $^{128-133}$Te isotopes without any restriction.

\subsection{Analysis of spectra}
The comparisons of calculated and experimental spectra for $^{124-127}$Te isotopes are shown in Figs. 1, 3, 5, 7.

\subsubsection{{\bf $^{124}_{~52}$Te$_{72}$:}\label{Te124}}

The spin sequence of the calculated positive parity energy levels is the same as in the experiment, 
however the energy levels $2^+_1$, $4^+_1$, $6^+_1$,
 $8^+_1$, $10^+_1$, $12^+_1$, $14^+_1$, $14^+_2$, $15^+_1$, and $16^+_1$ are  
124, 316, 451, 939, 1108, 1284, 1410, 1373, 1338, and 1594 keV lower than 
in the experiment, respectively. The calculated energy levels in the shell model are compressed as compared to the
experimental ones. This is because of the truncation while filling the neutrons in the model space, which we discuss in the details of calculation.
In the case of negative parity energy levels,  model predicts the $7^-$ (847 keV) level lower than in the 
experiment. The calculated $11^-$ level is 105 keV lower than the $9_2^-$.
The order of the calculated negative energy levels are the same as in the experiment.
 There is 586 keV energy difference between the levels $9^-_1$ and $9^-_2$ in shell model 
while this difference is very small (61 keV) in the experiment. The energy levels 
$9^-_1$, $9^-_2$, $11^-$, and $12^-$ are 981, 456, 1198 and 980 keV lower,
than in the experiment, respectively.  The r.m.s. deviation of the calculated positive parity states from the experimental ones is 1056 keV, while for negative parity states it 
 is 925 keV.

\begin{figure*}
\begin{center}
\includegraphics{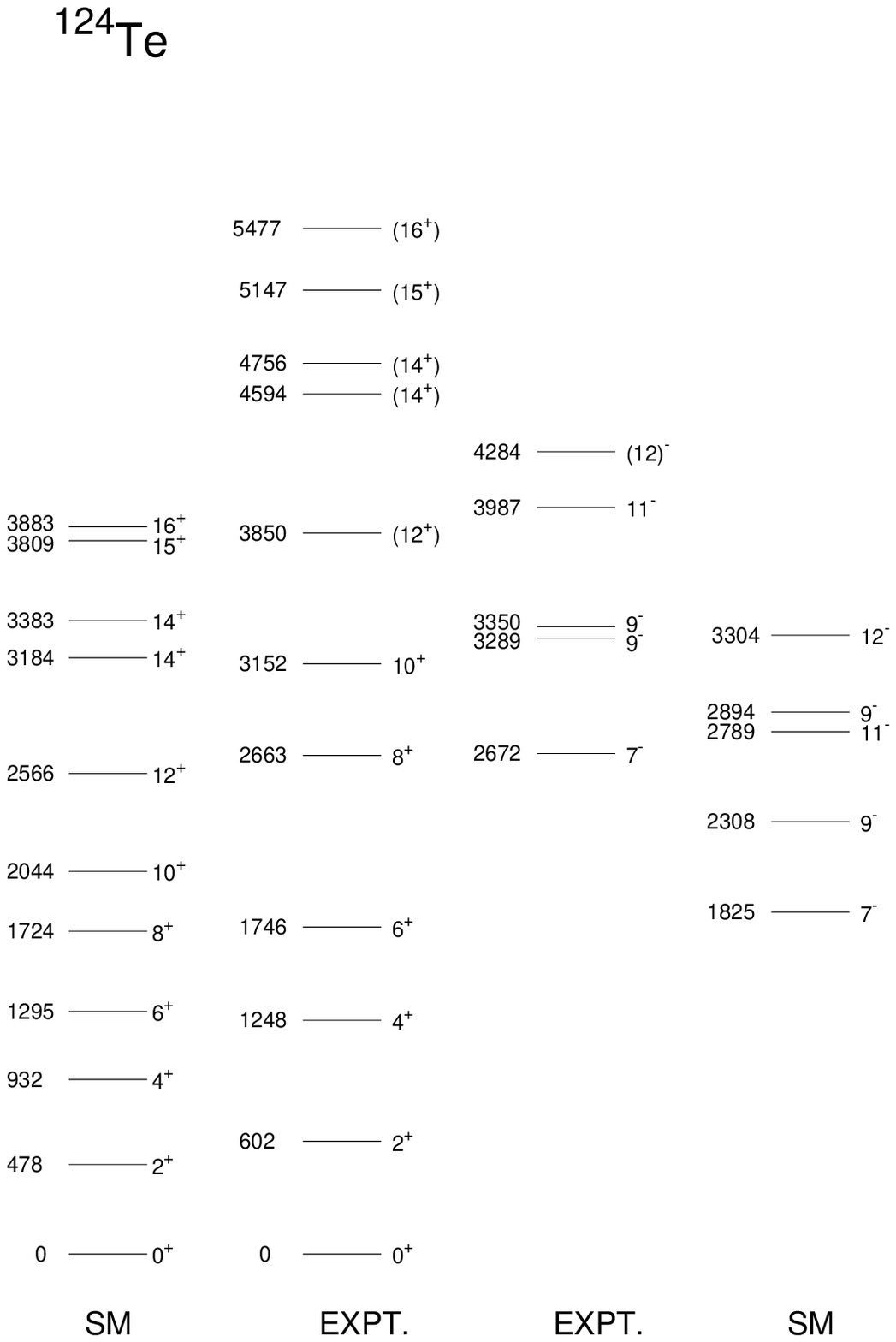}
\end{center}
\caption{
Comparison of calculated and experimental excitation spectra for $^{124}$Te 
using SN100PN interaction.}
\label{f_xe136}
\end{figure*}

From the analysis of the wave functions it is possible to identify which nucleon pairs are broken to obtain the total angular
momentum of the calculated states. The two components for neutrons and protons are $I_{n}$ and $I_{p}$, respectively. These
components are coupled to give the total angular momentum of each states. In the figs. 2 (a) -(d), we have shown results of positive parity 
states of $^{124}$Te.  The dominant component (46\%) of the 
$10^+$ state predicted at 2044 keV comes from  
the neutron pair breaking ($I_{n}$ = 10), the two protons being paired ($I_{p}$=0). The $12^+$ and $14_1^+$ states are due to breaking of 
both neutrons and protons pairs. The states $12^+$ and $14_1^+$ are collective states. The $14_2^+$ state calculated at 3383 keV has mainly $I_{p}$=6 (with  $I_{n}$=8-12), i.e., 
the proton pair being broken and the two angular momenta being fully aligned.
 The above three families are drawn with three different colors, the magenta color is for breaking
of neutron pairs, the green color is for that of protons and blue color is for many components with various values of $I_{n}$ and $I_{p}$.

The negative parity states are shown in figs. 2 (e) -(h). The dominant component (49 \%) of $7^-$ comes from the neutron pair breaking 
($I_{n}$ = 7), the two protons being paired ($I_{p}$=0). Similarly the $13^-$ (at 3247 keV) and $15^-$ (at 3717 keV) are also from breaking the
neutron pairs of $I_{n}$=13 and 15, respectively. Here, the two protons are paired ($I_{n}$=0).
The $11^-$ (at 2789 keV) shows many components. Thus, unlike the positive parity states, the negative parity states are not coming from the proton pair breaking.

\begin{figure*}
\resizebox{1.0\textwidth}{!}{
\includegraphics{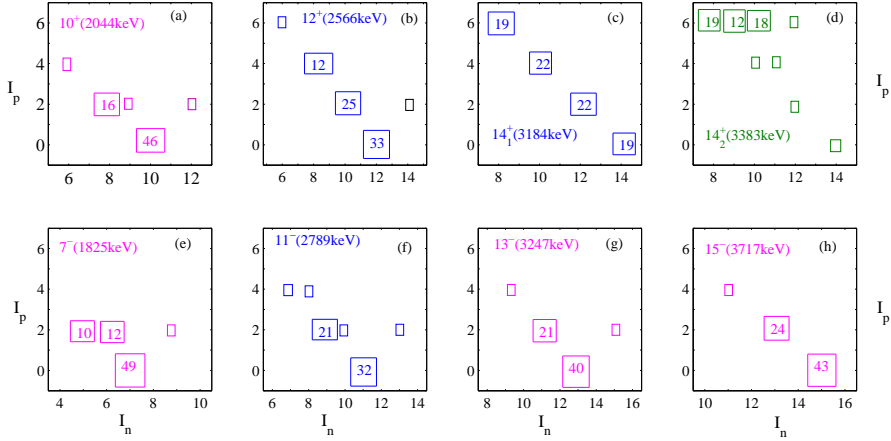}}
\caption{
Decomposition of the total angular momentum of selected states of $^{124}$Te  into their $I_n \otimes I_p$ components. The percentage above 10\%
 are written inside the squares, drawn with an area proportional to it. Percentage below 5\% are not written.}
\label{f_124Tepart}
\end{figure*}

\subsubsection{{\bf $^{125}_{~52}$Te$_{73}$:}\label{Te125}}
The $1/2^+$ is a ground state, while the isomeric $11/2^-$ state is at 145 keV in the experiment.
For this isotope shell model fails to predict the ground state correctly. It is $11/2^-$ in the calculation
and the next level is $3/2^+$ at 129 keV.
The $3/2^+$ energy level in the shell model calculation is higher by 94 keV than
in the experiment. After these two positive parity levels the order of the calculated positive parity energy levels are 
the same with the experiment, but the calculated values of the levels beyond $7/2^+$ are lower than in the experiment.
The calculated $7/2^+$ level is higher by 49 keV than in the experiment. 
The calculated energy levels $11/2^+$, $15/2^+$, $23/2^+$, $27/2^+$ and $31/2^+$ are 59,
143, 680, 778, and 751 keV lower than in the experiment,  respectively.
The calculated $11/2^-$ level is 145 keV lower than in the experiment.
The sequence of experimental $25/2^-_1$-$25/2^-_2$-$27/2^-_1$ levels is predicted as  $25/2^-_1$-$27/2^-_1$-$25/2^-_2$
in shell model. There is very small energy difference (16 keV) between 
$25/2^-_1$ and $27/2^-_1$ in the shell model, while this 
energy difference is 703 keV in the experiment.
Overall the spin sequence of the calculated negative energy levels is
the same with the experimental data.  The energy difference between the calculated and the
experimental levels is greater at higher spins.
 For the positive parity states, the r.m.s. deviation is 459 keV, while for the negative parity states 
it is 929 keV.

\begin{figure*}
\begin{center}
\includegraphics{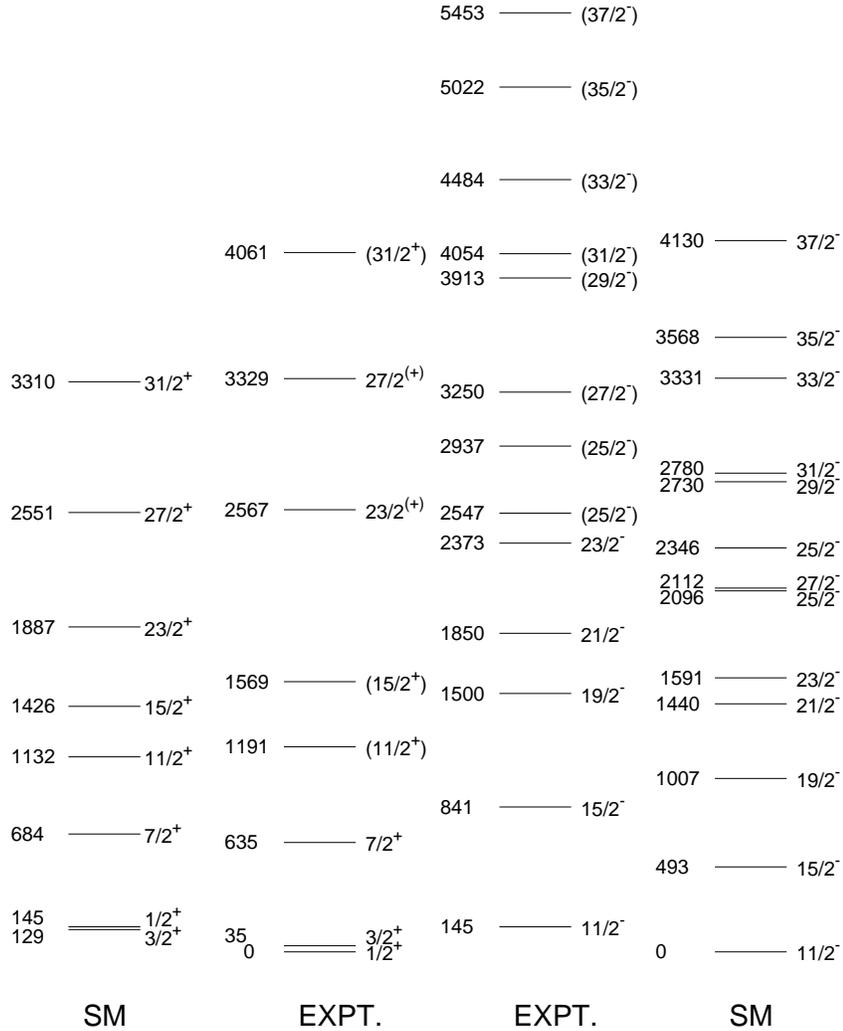}
\end{center}
\caption{
Comparison of calculated and experimental excitation spectra for $^{125}$Te 
using SN100PN interaction.}
\label{f_xe136}
\end{figure*}

In figs. 4 (a) -(d) the components $I_n$ an $I_p$ for negative parity states of $^{125}$Te are given. The $21/2^-$ state is predicted
at 1440 keV, comes from the proton pair breaking ($I_p = 6$), this has one odd neutron in the $h_{11/2}$ orbit.
The major components (41\%) of the  $27/2^-$ (2112 keV) state come from the neutron pair breaking ($I_n =27/2$), the two 
protons being paired ($I_p = 0$). The  $29/2^-$ (2730 keV) state comes from the proton  breaking ($I_p = 6$). 
The $31/2^-$ (2780 keV) state displays many components with various values of $I_n$ and $I_p$. 
The components of positive parity states are shown in figs. 4 (e) -(h). The $15/2^+$ state predicted at 1426 keV comes from 
the neutron breaking ($I_n =15 /2$), the two protons being paired ($I_p =0$).
The major component of $23/2^+$ increases to 51\% with the neutron pair breaking ($I_n = 23/2 $).
The $27/2^+$ state exhibits many components with various values of both $I_n$ and $I_p$ (similar results are obtained for
$31/2^+$ state at 3310 keV).

\begin{figure*}
\resizebox{1.0\textwidth}{!}{
\includegraphics{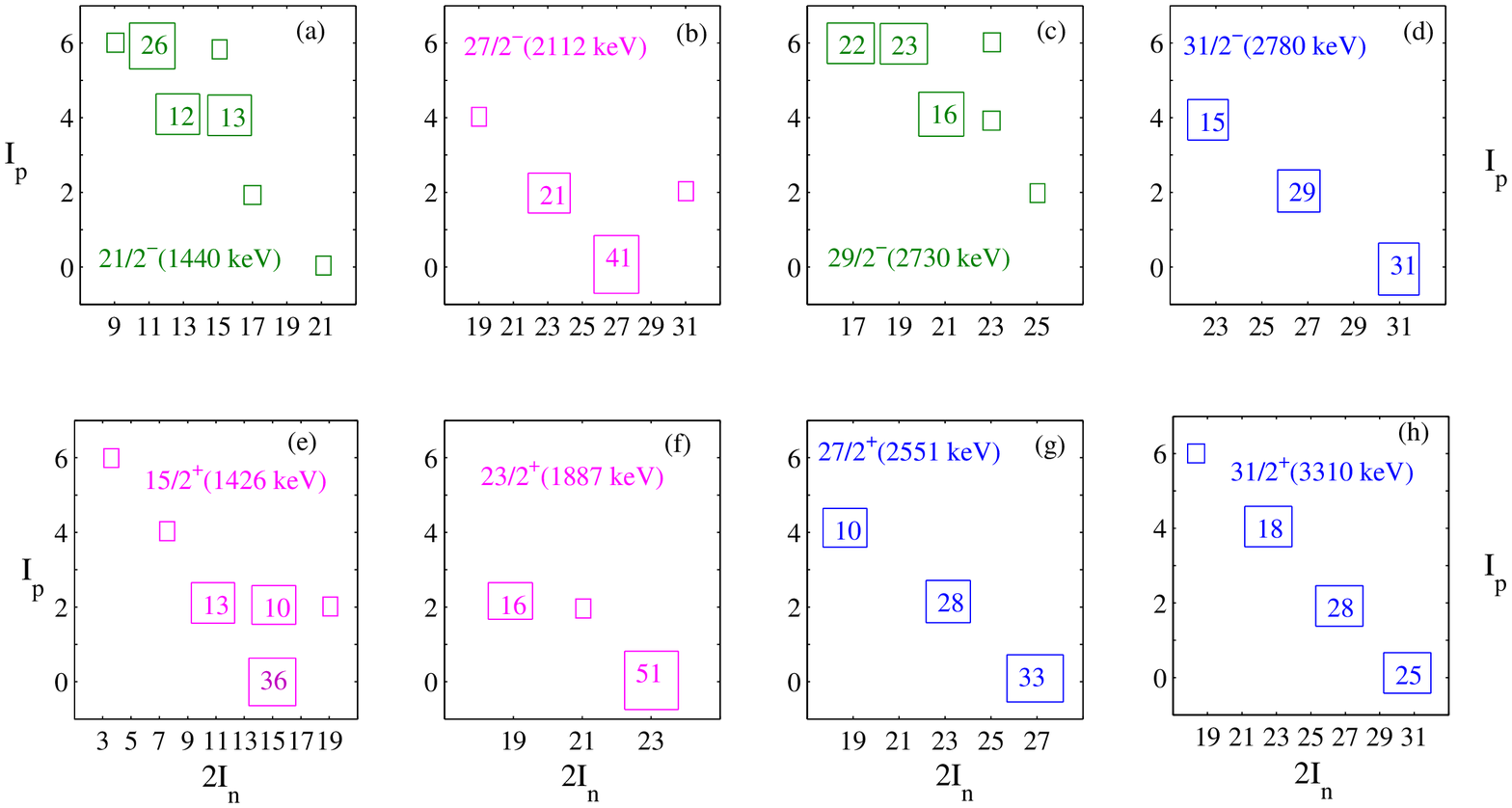}}
\caption{
Decomposition of the total angular momentum of selected states of $^{125}$Te  into their $I_n \otimes I_p$ components. The percentage above 10\%
 are written inside the squares, drawn with an area proportional to it. Percentage below 5\% are not written.}
\label{f_125Tepart}
\end{figure*}

\subsubsection{{\bf $^{126}_{~52}$Te$_{74}$:}\label{Te126}}

The calculated $2^+$, $4^+$ and $6^+$ levels are  118, 254 and 213 keV lower than in the experiment.
 The spin sequence of the positive parity states is the same as in the experiment, excluding 4137 keV
 experimental level, to which no spin and parity has been assigned yet.
 The calculated 
energy levels $8^+_1$, $10^+_1$, $12^+_1$, $13^+_1$, $14^+_1$, $14^+_2$, $15^+_1$, 
$16^+_1$, and $16^+_2$ are 823, 821, 925, 993, 1021, 953, 959, 1229 and 1254 keV lower
than in the experiment, respectively. The unassigned spin and the parity of the energy levels
 at 4137 keV cannot definitely be assigned with the shell model because of the large energy difference
 between calculated and experimental levels.

\begin{figure*}
\begin{center}
\includegraphics{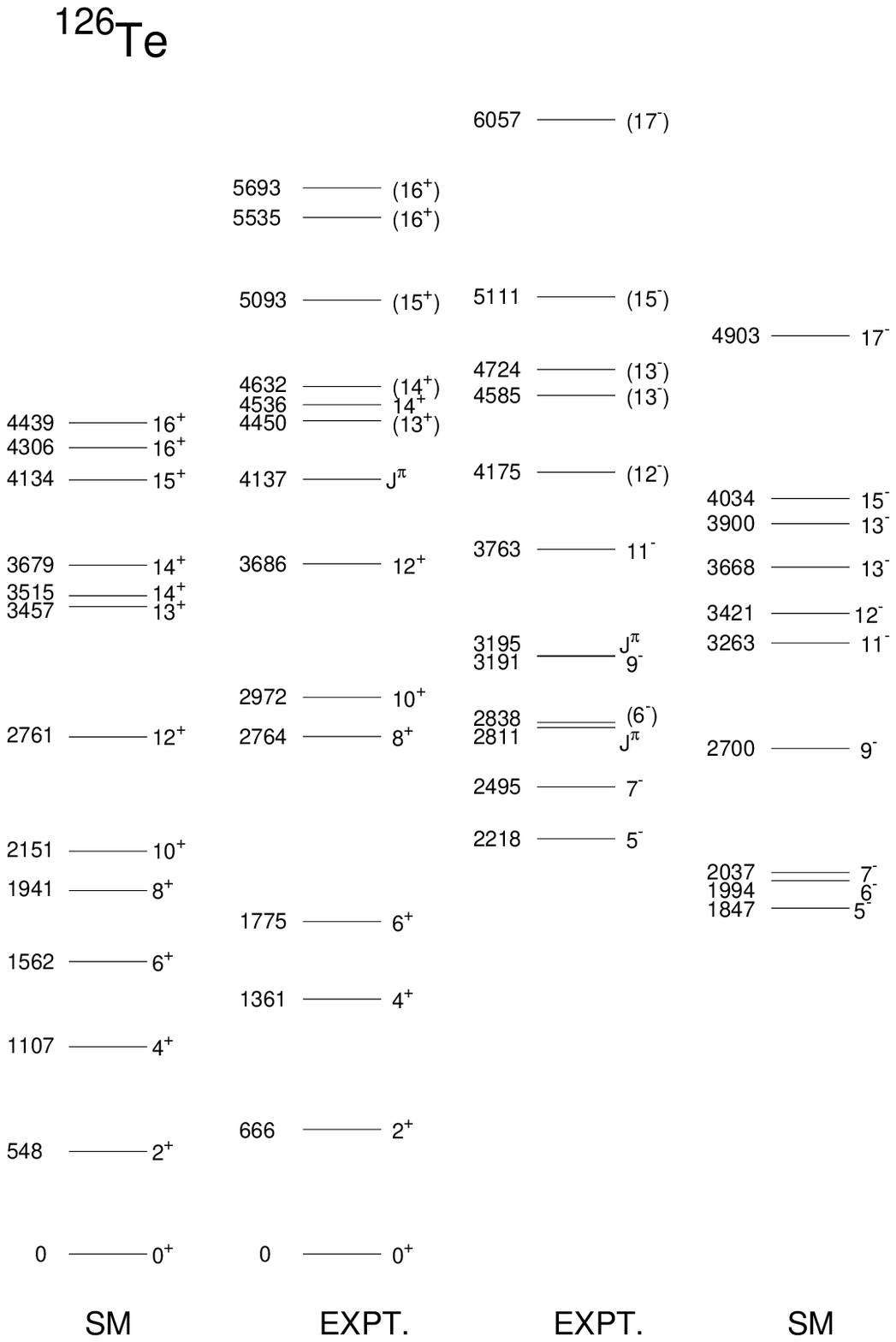}
\end{center}
\caption{
Comparison of  calculated and experimental excitation spectra for $^{126}$Te 
using SN100PN interaction.}
\label{f_xe136}
\end{figure*}

  The lowest negative parity energy level is $5^-$,  both in the calculation and the experiment. The calculated $5^-$ level 
is 371 keV lower than in the experiment. The energy levels $6^-$ and $7^-$
 are interchanged in shell model and the energy difference 
between these two pairs is 43 keV in shell model, while it is 343 keV in the experiment. 
The spin sequence of the calculated negative parity levels are the
 same as in the experiment, but energies are lower than in the experiment.
 In the shell model the energy difference between $11^-$ and $12^-$ levels are 158 keV 
while this difference is 412 keV in the experiment.
There is one unassigned level at 2811 keV between the $5^-$ and $6^-$ in the experiment. Shell model predicts this level at
1915 keV ($4^-$) which is not shown in fig. 5 because of the small difference in the energies. 
 The energy difference between the pair of $5^-$ and $6^-$ levels is
 147 keV in shell model, while this difference is 620 keV in the experiment.
The levels $13^-_1$, $13^-_2$, $15^-_1$, and $17^-_1$, are  917, 824, 1077, and 
1154 keV lower  than in the experiment, respectively.
  The r.m.s. deviation of calculated positive parity states from experimental ones is 844 keV, while for the negative parity states 
it is 783 keV.

In figs. 6 (a) -(h) the components of positive and negative parity states of $^{126}$Te are given.
The $12^+$ state is due to the neutron pair breaking.
For this isotope, the $10^+$ (2151 keV) has major component (55\%) from the neutron pair breaking ($I_n=10$), 
the two protons being paired ($I_p=0$). Similar feature exhibits the $12^+$ state with major component (36\%) 
from the neutron pair breaking ($I_n=12$), the two protons being paired ($I_p=0$).
The $14_2^+$ state at 3679 keV has mainly from  $I_{p}$=6 (with  $I_{n}$=8-12). Here, also the two angular
momenta are fully aligned and the proton pair being broken. The $14_1^+$ state at 3515 keV have mixed components.

The negative parity states show the same trend as in $^{124}$Te. The major components for $7^-$ (2037 keV) , $13^-$ (3668 keV) and $15^-$ (4034 keV)
increase, while $11^-$ (3263 keV) state component becomes more mixed. From $^{126}$Te to $^{128}$Te, the same trend is reported in
ref. \cite{teepja}. This shows that the reasonable truncation  is used in the calculation.

\begin{figure*}
\resizebox{1.0\textwidth}{!}{
\includegraphics{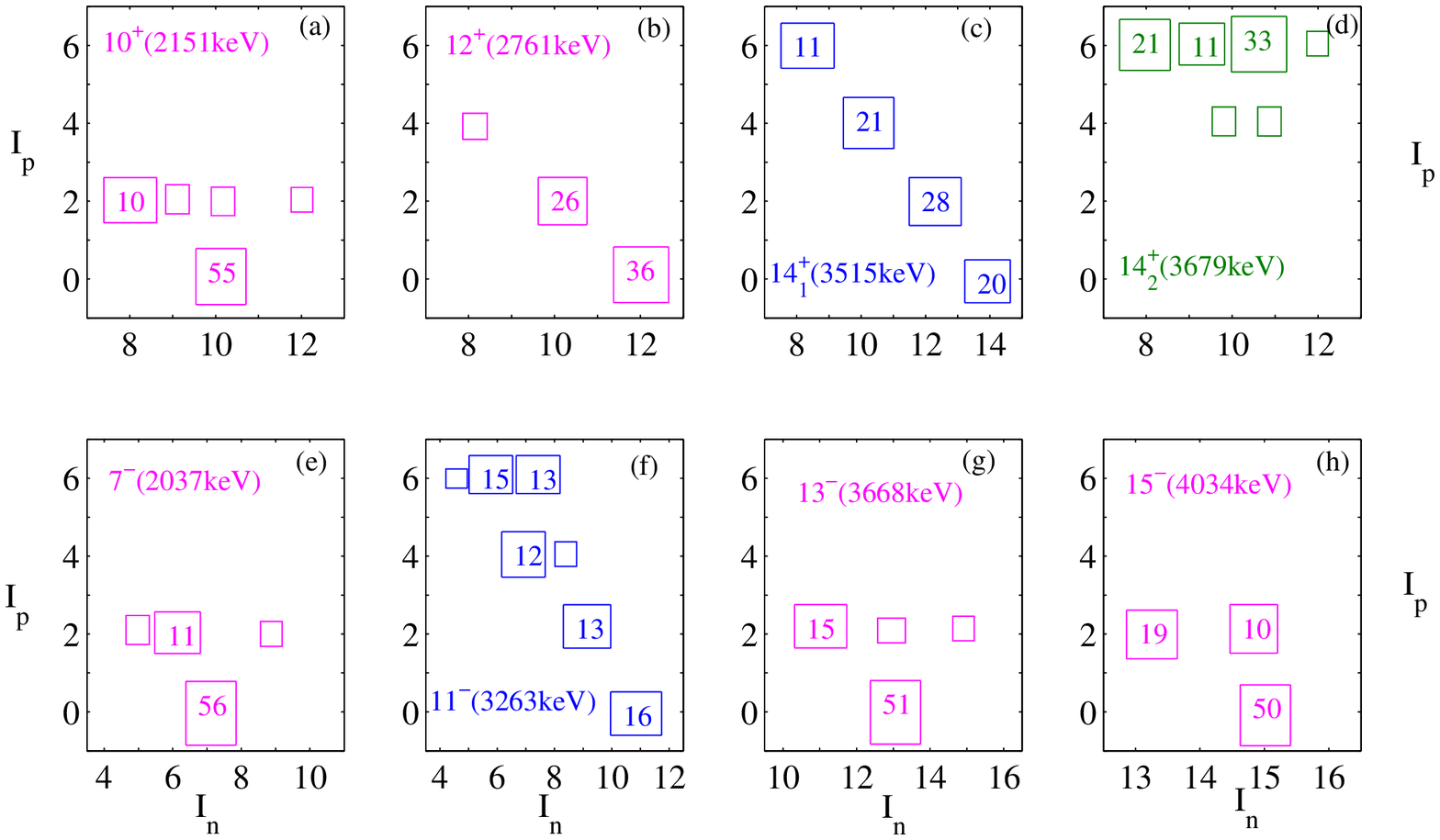}}
\caption{
Decomposition of the total angular momentum of selected states of $^{126}$Te  into their $I_n \otimes I_p$ components. The percentage above 10\%
 are written inside the squares, drawn with an area proportional to it. Percentage below 5\% are not written.}
\label{f_126Tepart}
\end{figure*}

\begin{figure*}
\begin{center}
\includegraphics{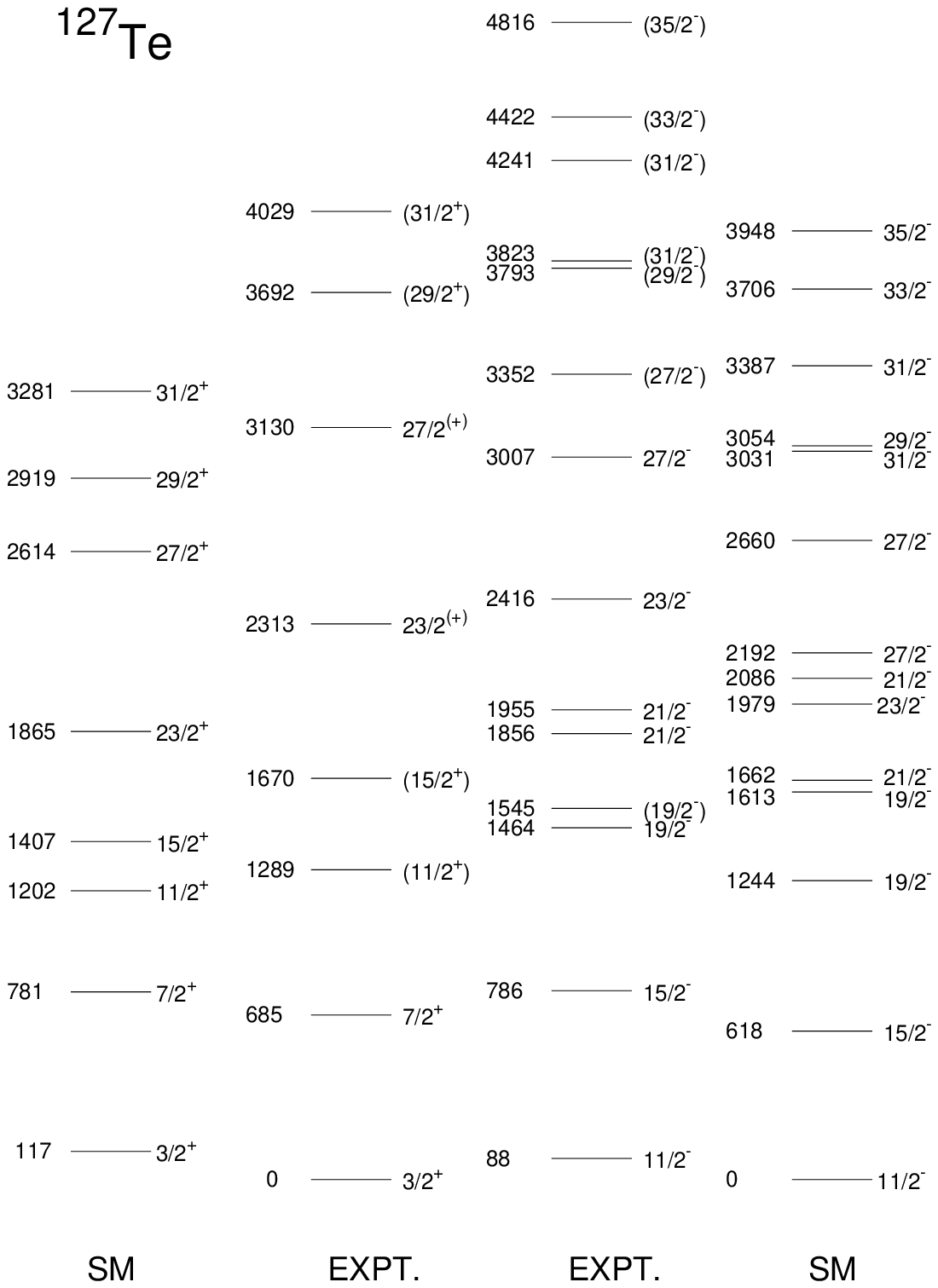}
\end{center}
\caption{
Comparison of calculated and experimental excitation spectra for $^{127}$Te 
using SN100PN interaction.}
\label{f_xe136}
\end{figure*}

\begin{figure*}
\resizebox{1.0\textwidth}{!}{
\hspace{0.1cm}\includegraphics{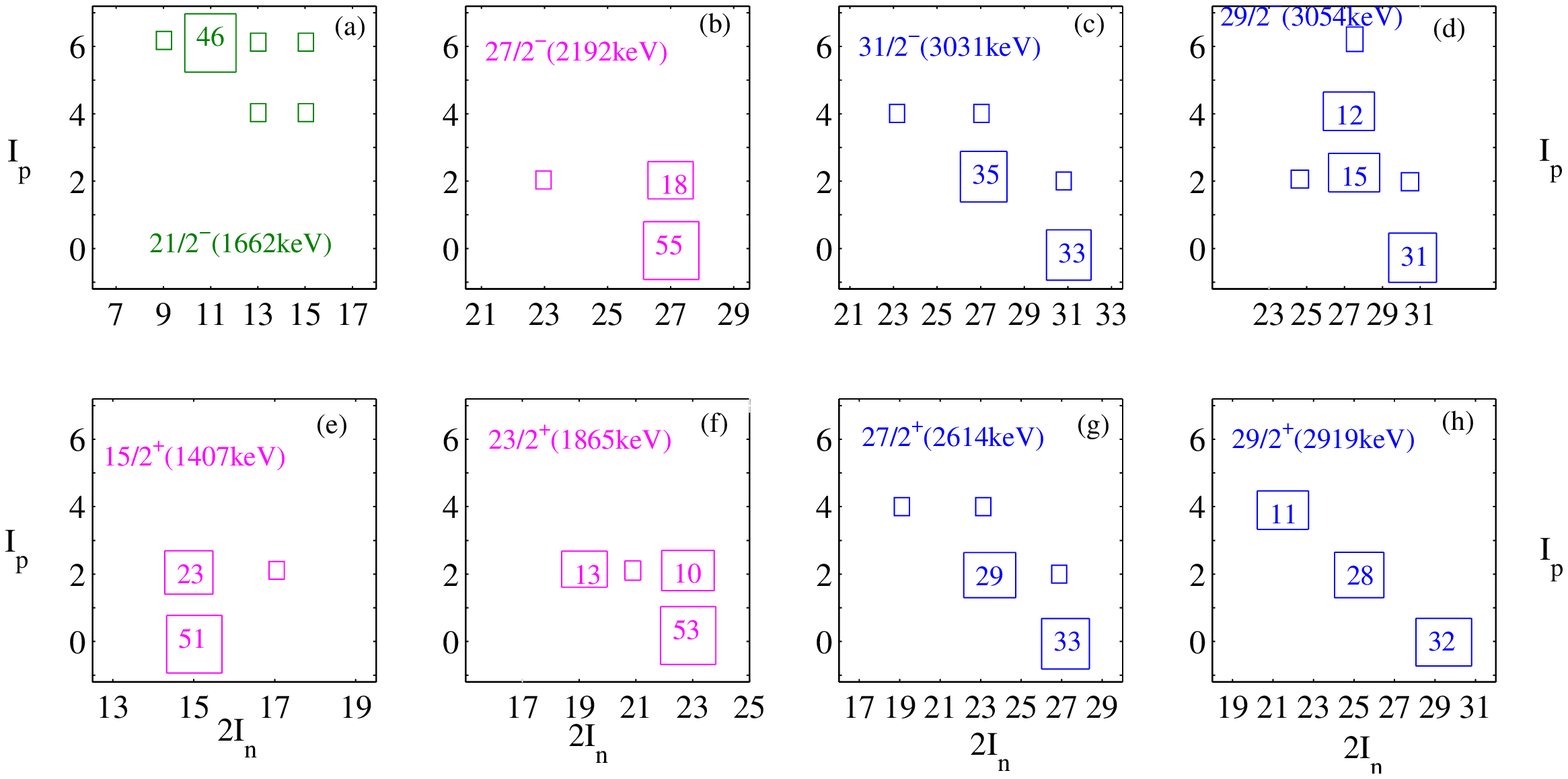}}
\caption{
Decomposition of the total angular momentum of selected states of $^{127}$Te  into their $I_n \otimes I_p$ components. The percentage above 10\%
 are written inside the squares, drawn with an area proportional to it. Percentage below 5\% are not written.}
\label{f_127Tepart}
\end{figure*}

\begin{table*}
\caption{Experimental and calculated $B(E2)$, $B(E3)$, $B(M2)$ and $B(M4)$ values in W.u. for different transitions. Experimental results are taken
fron ref. \cite{nndc,124,125,126,127,128,129,130,131,132,133}.}
\label{t_be2}
\begin{center}
\begin{tabular}{rcccc}
\hline
Nucleus  & ~~Transition & ~~Expt. &~~ Calc. I  &~~ Calc. II   \\   
    & ~~  & ~~ &$e_{\pi} =1.5e$,   & $e_{\pi} =1.5e$,  \\ 
    & ~~  & ~~ &$e_{\nu} =0.5e$   & $e_{\nu} =1.0e$   \\ 
\hline
$^{124}$Te &$B(E2;2^{+}_{1} \rightarrow 0^{+}_{1}$)\hspace{1.cm} & 31.1 (5) & 12.27 & 25.83\\ 
           &$B(E2;4^{+}_{1} \rightarrow 2^{+}_{1}$)\hspace{1.cm} & 97.529 (4)  & 18.27 & 37.22 \\
\hline
$^{125}$Te &$B(E2;3/2^{+}_{1} \rightarrow 1/2^{+}_{1}$)\hspace{1.cm} & 11.9(24) & 0.55 & 1.39 \\ 
            &$B(E2;7/2^{+}_{1} \rightarrow 3/2^{+}_{1}$)\hspace{1.cm} & 4.8(24) & 9.88  & 20.79 \\ 
\hline
$^{126}$Te &$B(E2;2^{+}_{1} \rightarrow 0^{+}_{1}$)\hspace{1.cm} & 25.4 (7)& 11.12  & 24.19 
\\ 
           &$B(E2;4^{+}_{1} \rightarrow 2^{+}_{1}$)\hspace{1.cm} & 34(16) & 16.52  & 34.97
   \\
           &$B(E2;6^{+}_{1} \rightarrow 4^{+}_{1}$)\hspace{1.cm} & 17.8(6) & 17.09  & 34.43 \\
            &$B(E2;10^{+}_{1} \rightarrow 8^{+}_{1}$)\hspace{1.cm} & 2.50(19) & 0.14  & 0.38 
 \\
\hline
$^{127}$Te &$B(E2;7/2^{+}_{1} \rightarrow 3/2^{+}_{1}$)\hspace{1.cm} & N/A & 5.40  & 11.01 \\  
                            
\hline
$^{128}$Te &$B(E2;2^{+}_{1} \rightarrow 0^{+}_{1}$)\hspace{1.cm} & 19.62(18) & 10.42   & 22.15 \\ 
           &$B(E2;6^{+}_{1} \rightarrow 4^{+}_{1}$)\hspace{1.cm} & 9.7(6) &  5.51  & 9.69  \\ 
            &$B(E2;10^{+}_{1} \rightarrow 8^{+}_{1}$)\hspace{1.cm} & 1.40(12) & 2.88  & 6.80 \\            
\hline
$^{130}$Te &$B(E2;2^{+}_{1} \rightarrow 0^{+}_{1}$)\hspace{1.cm} & 15.1(3) & 8.04  &  15.82\\ 
           &$B(E2;6^{+}_{1} \rightarrow 4^{+}_{1}$)\hspace{1.cm} & 6.1(3) & 3.63   & 5.79 \\ 
\hline
$^{131}$Te &$B(E3;23/2^{+}_{1} \rightarrow 19/2^{-}_{1}$)\hspace{1.cm} & 0.0151(20) &  0.0044& 0.018\\ 
           
\hline
$^{132}$Te &$B(E2;6^{+}_{1} \rightarrow 4^{+}_{1}$)\hspace{1.cm} & 3.3(2) & 2.61  & 3.49 \\ 
          
\hline
$^{133}$Te &$B(E2;19/2^{-}_{1} \rightarrow 15/2^{-}_{1}$)\hspace{1.cm} & 2.56(14)&  2.55 &2.51\\
\hline
\hline
Nucleus  & ~~Transition & ~~Expt. &~~ Calc. I  &~~ Calc. II    \\   
    & ~~  & ~~ & $g_s^{\rm eff} =g_s^{\rm free}$    & $g_s^{\rm eff} =0.7g_s^{\rm free}$  \\ 
           
\hline
$^{127}$Te &$B(M4;11/2^{-}_{1} \rightarrow 3/2^{+}_{1}$)\hspace{1.cm} & 3.6 &  16.21 &  7.95\\ 
\hline
$^{130}$Te &$B(M2;7^{-}_{1} \rightarrow 6^{+}_{1}$)\hspace{1.cm} & 0.013 (3) &  0.002& 0.00006 \\ 

\hline
$^{131}$Te &$B(M4;11/2^{-}_{1} \rightarrow 3/2^{+}_{1}$)\hspace{1.cm} & 4.59 &  21.75 & 10.65 \\ 
\hline
$^{133}$Te &$B(M4;11/2^{-}_{1} \rightarrow 3/2^{+}_{1}$)\hspace{1.cm} & 4.7(6) & 24.3 &  11.9\\ 
\hline           
\end{tabular}
\label{be2}
\end{center}
\end{table*}

\subsubsection{{\bf $^{127}_{~52}$Te$_{75}$}:\label{Te127}}

 The shell model predicts 
$3/2^+$ level at 117 keV which is the experimental ground state. The spin sequence of 
the positive parity energy levels beyond $7/2^+$ is exactly the same and the energies are lower than in the experiment. 
The $7/2^+$ level in shell model is higher by 96 keV than in the experiment. The tentative  $11/2^+$ level at 
1289 keV in the experiment is confirmed by shell model as a $11/2^+$ level. The $15/2^+$ level in shell
 model is lower by 263 keV than in the experiment.
The energy levels $23/2^+$, $27/2^+$, $29/2^+$, and $31/2^+$ are 448, 516
773, and 748 keV lower than in the experiment, respectively. 

Shell model predicts $11/2^-$ as the ground state which is the lowest negative parity 
energy level at 88 keV in the experiment. The order of the calculated energy levels
up to $21/2^-_1$ is the same as in the experiment, the $23/2^-$  and $21/2^-_2$ are interchanged in shell model. The $23/2^-$ level is 437 keV lower, and the 
$21/2^-_2$ is higher by 131 keV than in the experiment.
The energy difference between $19/2^-_1$ and $19/2^-_2$
is 369 keV in shell model, while this difference is 81 keV in the experiment.
The energy levels $27/2^-_1$, $27/2^-_2$,  $29/2^-_1$, $31/2^-_1$, $31/2^-_2$, $33/2^-$,
and $35/2^-$ are   815, 692, 792, 739, 854, 716 and 868 keV lower than in the experiment, 
respectively.
 The r.m.s. deviation of calculated positive parity states from experimental ones is 464 keV, while for the negative parity states 
 it is 576 keV. The r.m.s. deviation is decreasing as we move from  $^{124}$Te to  $^{127}$Te. This is because we are approaching towards the
 shell closure  and the truncation effect is minimizing in the energy level prediction.

In  figs. 8 (a) -(d) the components of negative parity states of $^{127}$Te are given. The $21/2^-$ state (46 \%) predicted
at 1662 keV comes from the breaking of the proton pair ($I_p = 6$), this has one odd neutron in the $h_{11/2}$ orbit.
The major components (55\%)  of the  $27/2^-$ (2192 keV) state come from the neutron pair breaking ($I_n =27/2$), the two 
protons being paired ($I_p = 0$). The $31/2^-$ (3031 keV) state exhibits many components with various values of both $I_n$ and $I_p$ (similar results are obtained for $29/2^-$ state at 3054 keV). 

The components of positive parity states are shown in figs. 8 (e) -(h). The $15/2^+$ state (51\%) predicted at 1407 keV comes from 
the neutron pair breaking ($I_n =15 /2$), the two protons being paired ($I_p =0$). Similarly, for $23/2^+$ state (53 \%) predicted at 1865 keV
comes from the neutron breaking ($I_n =23 /2$), the two protons being paired ($I_p =0$).
The $27/2^+$ state ( 2614 keV) exhibits many components with various values of both $I_n$ and $I_p$. Similar results are obtained for
$29/2^+$ state at 2919 keV. For the positive parity states trends are similar, as we move from $^{125}$Te to $^{127}$Te. However, in the case of 
$^{129}$Te, the $29/2^+$ state (3422 keV) is due to the proton pair breaking ($I_p =6$) \cite{teepja}, while this state in $^{127}$Te at 2919 keV has mixed components.

 Previously reported shell model results with the same interaction for nuclei around the $N=82$  shell closure
are in very good agreement with the experimental data on the 100 keV level \cite{JPGN82,PhysRevC.71.044317}. However, present shell model results for $^{124-127}$Te
isotopes show large discrepancies - up to  500 and even 1000 keV, as compared to the experiment.
It is very difficult to estimate correctly the predictive power of this interaction  in the present case, because of the truncation employed in $^{124-127}$Te isotopes.
Here, both proton and neutron orbitals are open, thus dimensions become huge (for example it is around 820 millions
in m-scheme for $^{124}$Te). Thus, to perform feasible shell model calculations we have employed truncation for the neutrons orbital.
The discrepancies between the calculated and experimental values may be caused by this.

\section{Transition probability analysis}

The comparison of the transition probabilities with the experiment
data 
\cite{124,125,126,127,128,129,130,131,132,133} is given in Table~\ref{t_be2}.
The choice of proton/neutron effective charges in Sn region is very important. For the nuclei where both proton and neutron shells are in
82-126 region, it was claimed that for B(E2;$2^{+}_{1} \rightarrow 0^{+}_{1}$) transition in $^{138}$Ba, corresponding to 
the experimental 460(18) $e^2fm^4$ value, the two popular interactions SMPN  and CWG  predict better results corresponding to
$e_p\simeq 1.4e$ and $e_n\simeq0.7e$ \cite{ss2}, while in the study for Sn isotopes, the experimental 
B(E2;$2^{+}_{1} \rightarrow 0^{+}_{1}$) values for $^{124}$Sn to $^{130}$Sn isotopes were studied with three set of neutron effective charges,
$e_n=0.70e$, $e_n=0.85 e$ and $e_n=1.0e$ (see, fig. 11 (c), in ref. \cite{Lozeva}). It was observed that $e_n=0.85e$ best describes the experimental value.
But the variation of 0.70$e$ at $N=82$ to 1.00$e$ at $N\sim70$ might be more appropriate as we go from $^{130}$Sn to $^{124}$Sn.
Recently, an effective neutron charge 1.0e for $^{100}$Sn core is reported in ref. \cite{thesis,110Sn,PRL2} to correctly reproduce $B(E2)$ values for 
Sn isotopes with $A=102-130$.

In view of above findings, we performed calculation with the two set of effective charges (1.5$e$, 0.5$e$) and (1.5$e$, 1.0$e$) in the present work.
The  overall calculated values of B(E2;$2^{+}_{1} \rightarrow 0^{+}_{1}$) transition
probabilities are in good agreement with the experimental ones with $e_{\pi} =1.5e$ and $e_{\nu} =1.0e$. 
Thus, the results with the set of effective charges (1.5$e$, 0.5$e$) and (1.5$e$, 1.0$e$) indicate strong sensitivity of $E2$ transition probabilities to the effective charge of neutrons.
The $B(E2;6^{+}_{1} \rightarrow 4^{+}_{1}$) values for $^{126,128,130,132}$Te are in good agreement with the experimental value, while
$B(E2;10^{+}_{1} \rightarrow 8^{+}_{1}$) for $^{126}$Te is slightly less and for $^{128}$Te is larger than in the experiment.
The calculated $B(M4;11/2^{-}_{1} \rightarrow 3/2^{+}_{1}$) for $^{127}$Te are 16.21 W.u. ( $g_s^{\rm eff} =g_s^{\rm free}$ ) and 7.95 W.u. ( $g_s^{\rm eff} = 0.7 g_s^{\rm free}$ ), corresponding experimental
value is 3.6 W.u. For $^{130}$Te the calculated $B(M2;7^{-}_{1} \rightarrow 6^{+}_{1}$) value is 0.00139 W.u. ( $g_s^{\rm eff} =g_s^{\rm free}$ ), while the corresponding experimental
value is 0.013(3) W.u.  We have also calculated  $B(E2;7/2^{+}_{1} \rightarrow 3/2^{+}_{1}$) value for $^{127}$Te, however,  experimental data
for this transition is not available.  The r.m.s. deviation for 
$B(E2;0^{+}_{1} \rightarrow 2^{+}_{1}$) transitions is 0.049 $e^2b^2$ (with $e_p=1.5e$ and $e_n=1.0e$)
and 0.225 $e^2b^2$ (with $e_p=1.5e$ and $e_n=0.5e$).

\begin{figure}[htp]
\begin{center}
\resizebox{0.9\textwidth}{!}{
\includegraphics{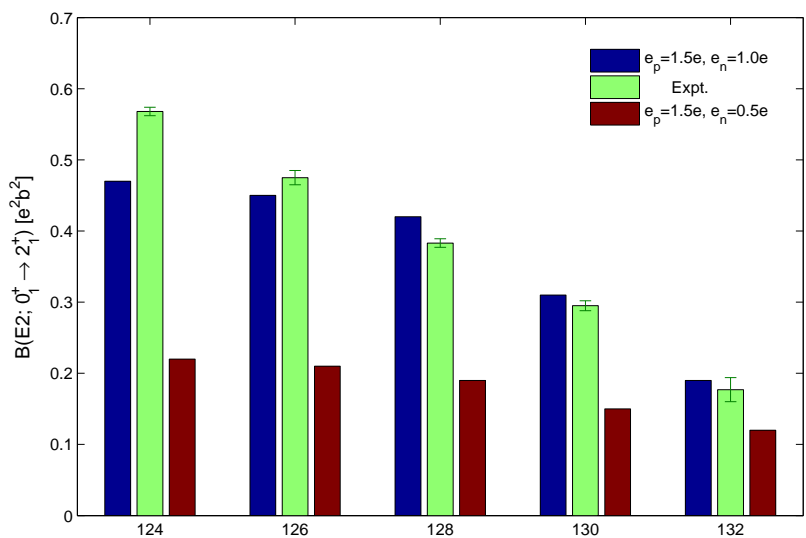}}
\end{center}
\caption{
Calculated and experimental value of $B(E2;0^{+}_{1} \rightarrow 2^{+}_{1})$ for even Te isotopes  [$A=124-132$].}
\end{figure}

 In recent past many experimental affords have been done to measure the $B(E2;0^{+}_{1} \rightarrow 2^{+}_{1}$) for Sn isotopes beyond  $^{108}$Sn   
\cite{110Sn,PRL2}. These results indicate the transition rates are almost independent of mass number $A$.  The experimentally observed asymmetric behavior
of the $B(E2)$ values with respect to midshell ($N=66$) is in disagreement with large-scale shell model calculations.
The systematic theoretical study of the transition rates of Sn isotopes with two model spaces one with $^{90}$Zr core and another with $^{100}$Sn
core using the same effective interaction is shown in  ref. \cite{110Sn}.
These results indicate that
the core polarization due to particle-hole excitations is dominating for the lighter Sn isotopes.  In this work, it was observed that the standard 
set of effective charges {\it i.e.} $e_p=1.5e$ and $e_n=0.5e$ predict good agreement with the experimental data for $^{90}$Zr core with $4p-4h$ excitations, while if we take $^{100}$Sn as a core, then to reproduce
 correct experimental trend, we should increase effective charges for neutrons from $e_n=0.5e$ to $e_n=1.0e$. 
 The experimental $B(E2;0^{+}_{1} \rightarrow 2^{+}_{1})$ values for Te isotopes show the same trend \cite{raman,back,saxena,doncel}. Thus, to get the correct $B(E2)$ values from shell model, we should take into account core polarization effects with standard  set of effective charges, although it is very difficult to perform full-fledged shell
 model calculation for Te isotopes around mid-shell. In our case,  since we put truncation, and thus increase of the neutron effective charge is obvious.
 Our calculated results indicate that the $^{100}$Sn core is not robust for Te isotopes, thus core-polarization effects 
 may be needed and further improvement in the effective interaction (with $^{100}$Sn core) is required. Also further experimental investigation
 is needed to measure $B(E2;0^{+}_{1} \rightarrow 2^{+}_{1})$ transition in lighter isotopes
 of Sn, Te and even beyond.  The calculated and experimental $B(E2)$'s values are shown in fig. 9. These results indicate that for $^{128,130,132}$Te,
 the neutron effective charge $e_n=1.0e$ gives good agreement with experimental data, while for $^{128,130,132}$Te small deviation
is due to the truncation used in our calculation.

\section{Estimate for contributions from three-body forces }

Inclusion of 3N force in shell-model calculation leads to the shift of all energies for a nucleus.
It may also shift the individual states different way. The importance of three-body forces for lighter nuclei were
reported in literature \cite{quesne,polls,zele,poves,piet}. The average energy shift of states with n valence particles due to a 
3N force can be written as  

\begin{equation}
\begin{array}{l}
\lefteqn{\Delta E_3 (n) } \hspace{15mm} = ~  {n \choose 3}  {\bar V_3 }
\end{array}
\end{equation}
where ${n \choose 3}$ is the binomial coefficient and $\bar V_3$ is the average matrix element which is given by, 
\begin{equation}
\begin{array}{l}
\lefteqn{\bar V_3 } \hspace{5mm} = ~\frac{\sum\limits_{\alpha \beta \gamma JT} (2J+1)(2T+1){\left\langle \alpha \beta \gamma |V_3|\alpha \beta \gamma \right\rangle _{JT}}}{\sum\limits _{\alpha \beta \gamma JT} {\left\langle \alpha \beta \gamma | \alpha \beta \gamma \right\rangle _{JT}(2J+1)(2T+1)}}.\\
\end{array}
\end{equation}

In the case of $^{124-127}$Te, we can estimate contribution of three-body forces from eqs. (1) and (2) by considering occupancy of
neutron $h_{11/2}$ orbital. As in the reference \cite{quesne},
it was shown that the interactions of order higher than 2 are important in the $0f_{7/2}$ shell for $fp$ shell nuclei. In our case for Te isotopes,
with 2 protons outside $Z=50$ core, there is no contribution of three-body forces on $p-p$ part of the SN100PN interaction.
For $^{124,126}$Te, with even $J$ values from $0^+$-$16^+$, the
occupancy of neutron $h_{11/2}$ orbital is varying from $\sim$  6 ($^{124}$Te) to $\sim$ 8 ($^{126}$Te) as shown in fig. 10. Thus, as we move from $^{124}$Te to $^{126}$Te
the effect of 3N forces will change from ${6 \choose 3}$$\bar V_3$ to ${8 \choose 3}$$\bar V_3$. From this we can conclude that 
the effect of 3N forces is 2.8 times larger for $^{126}$Te in comparison to $^{124}$Te. Although this effect is constant for $0^+$-$16^+$,
because occupancy is almost the same for $h_{11/2}$ orbital for these states. 

\begin{figure}[htp]
\begin{center}
\resizebox{0.6\textwidth}{!}{
\includegraphics{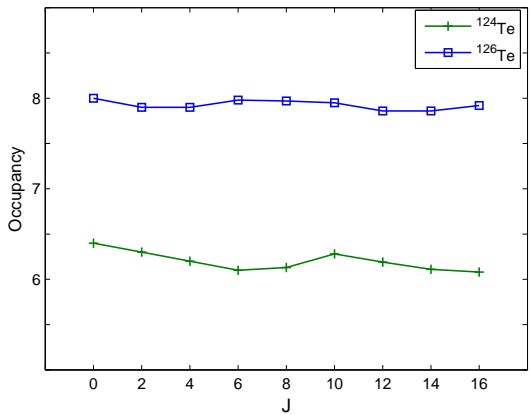}}
\end{center}
\caption{
Variation of occupancy for $h_{11/2}$ orbital with spins in $^{124}$Te and $^{126}$Te isotopes.}
\end{figure}

\section{Results with modified Hamiltonian}
It is believed that once neutrons start filling the $h_{11/2}$ orbital, then $n-n$ part of the effective interaction becomes too attractive. Thus, we decided to tune affect of $h_{11/2}$
orbital on $d_{3/2}$ and $s_{1/2}$ orbitals only (here we have
ignored its impact on $g_{7/2}$ and $d_{5/2}$ orbitals).
For this, we have changed only those two-body matrix elements corresponding to
these combinations. In total, we have modified 18 set of two-body matrix elements by adding +0.200 MeV in the original $n-n$ part.
Results for low-lying states change significantly with this modification. We have shown the shell model results corresponding to this modification in fig. 11 ($^{124,126}$Te) 
and fig. 12 ($^{125,127}$Te). The difference between calculated and experimental first two excited states ($2_1^+$ and $4_1^+$) in $^{124,126}$Te are within difference of $\sim$ 100 keV.
In the case of $^{125,127}$Te, the difference for $7/2_1^+$ and $11/2_1^+$ states are within difference of $\sim$ 200 keV.

\begin{sidewaysfigure}
\begin{center}
\includegraphics[width=13.5cm,height=14.6cm,keepaspectratio]{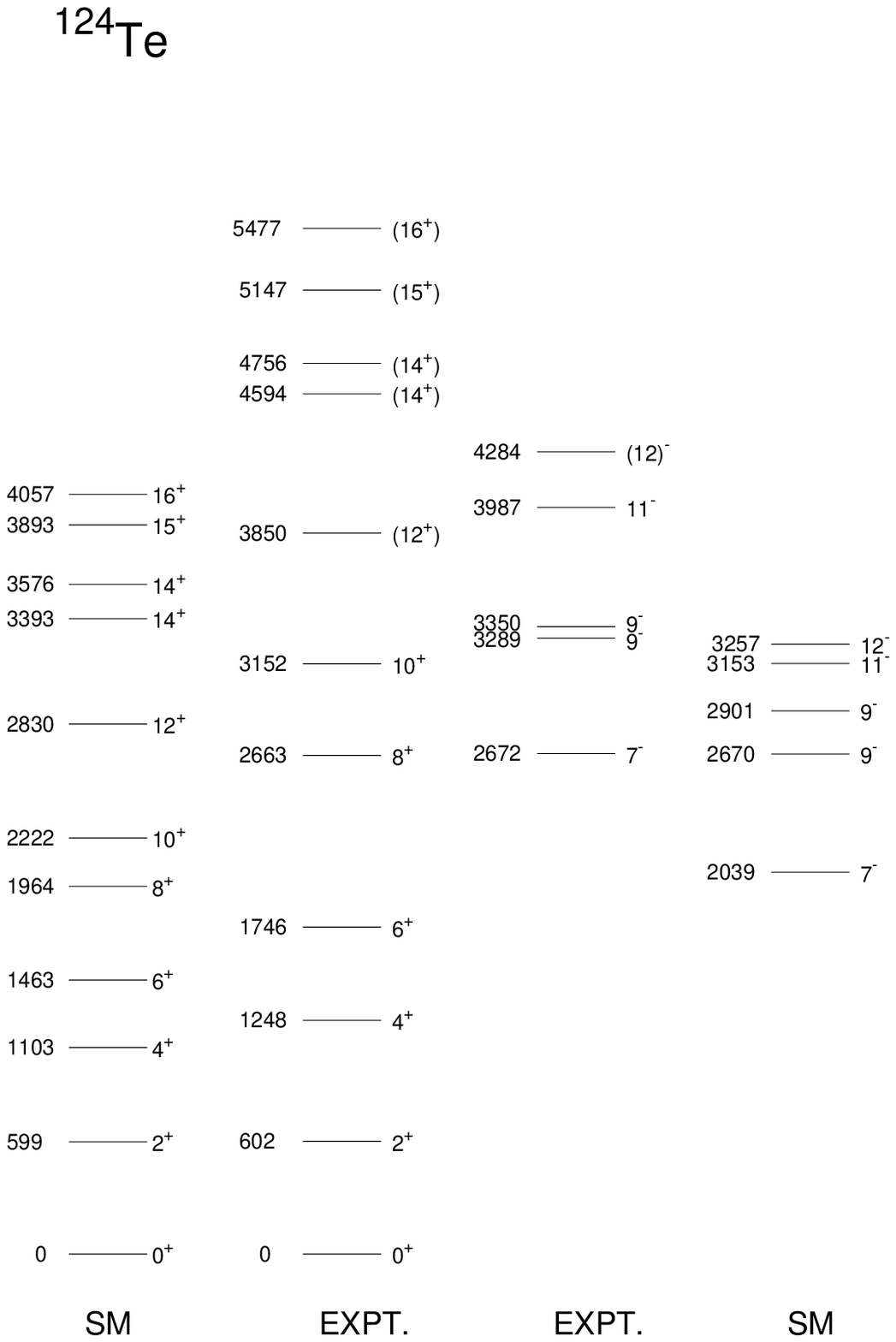}
\includegraphics[width=13.5cm,height=14.6cm,keepaspectratio]{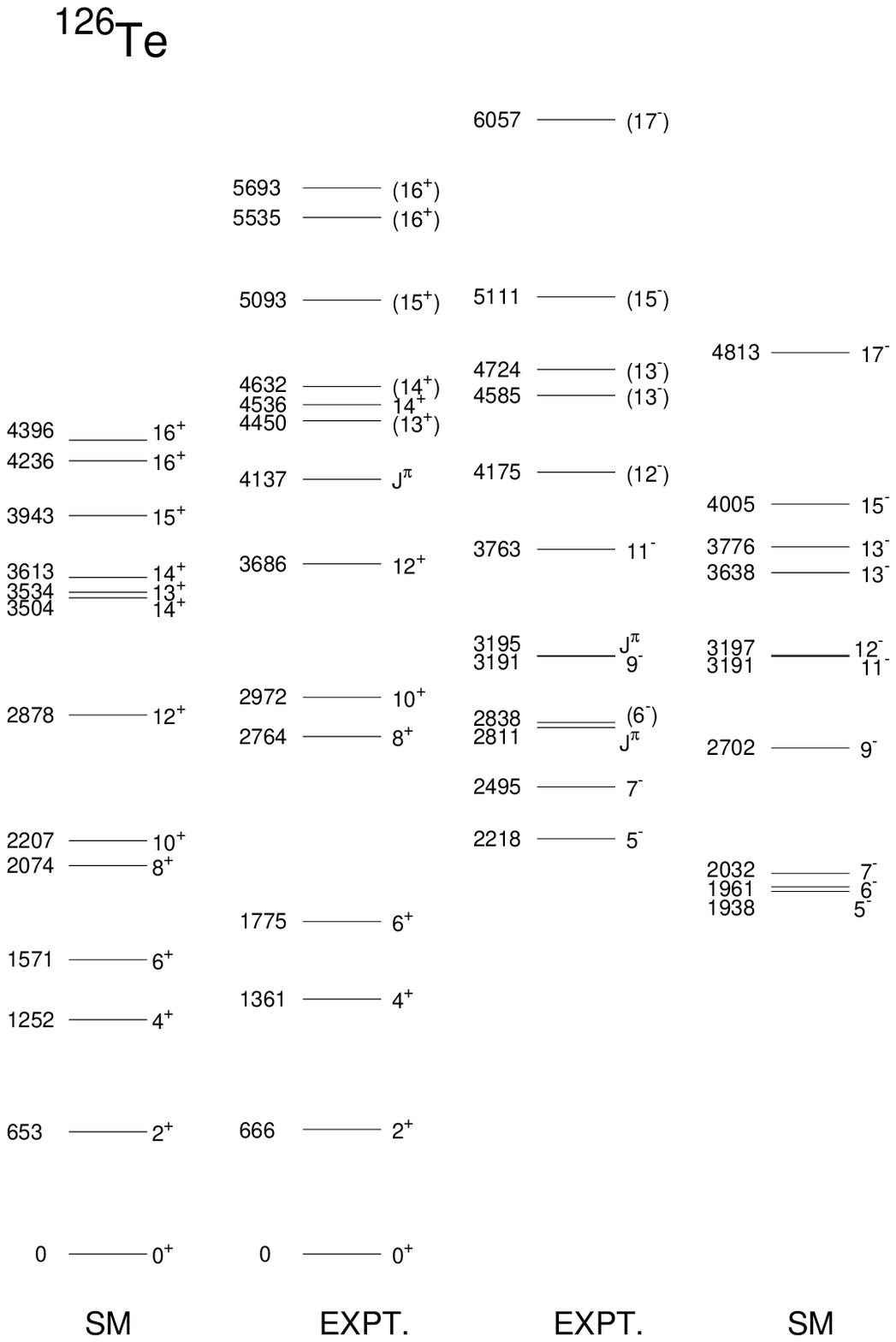}
\end{center}
\caption{
Comparison of calculated and experimental excitation spectra for $^{124,126}$Te 
using modified SN100PN interaction.}
\label{f_xe139}
\end{sidewaysfigure}

\begin{sidewaysfigure}
\begin{center}
\includegraphics[width=13.5cm,height=14.6cm,keepaspectratio]{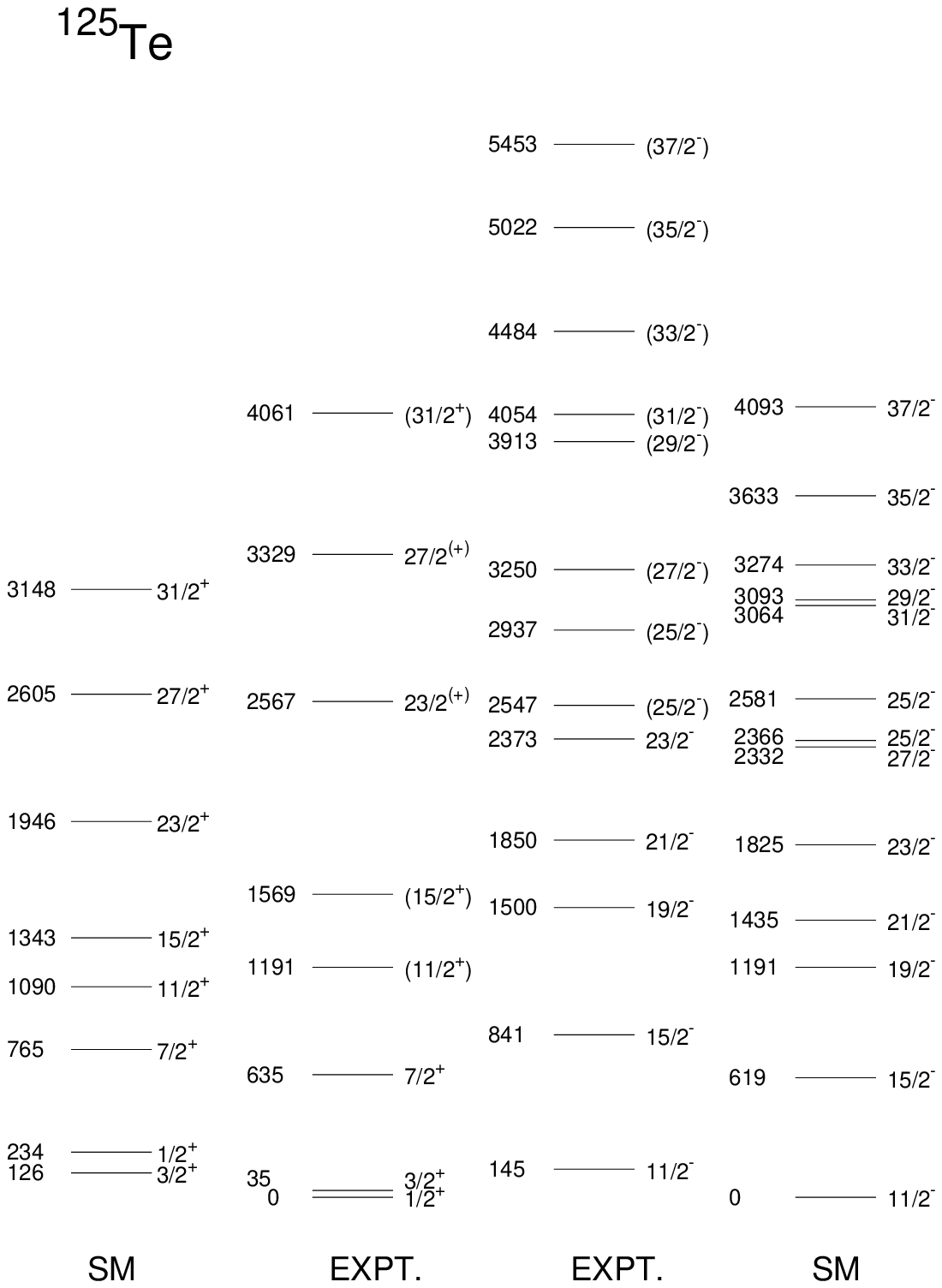}
\includegraphics[width=13.5cm,height=14.6cm,keepaspectratio]{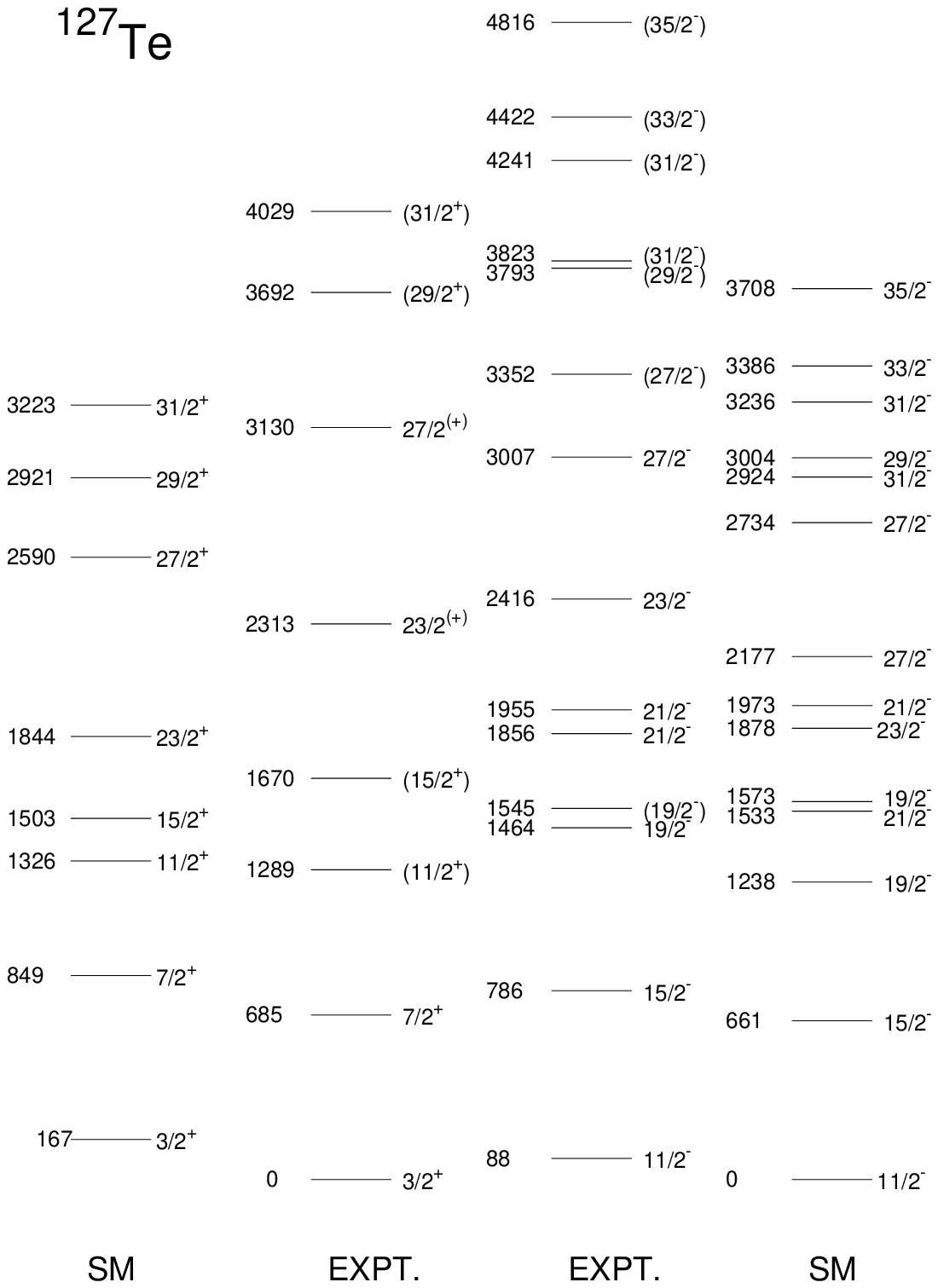}
\end{center}
\caption{
Comparison of calculated and experimental excitation spectra for $^{125,127}$Te 
using modified SN100PN interaction.}
\label{f_xe139}
\end{sidewaysfigure}

\section{Summary}
We have performed shell model calculation for the recently available experimental data for $^{124}$Te, $^{125}$Te, $^{126}$Te and $^{127}$Te isotopes.
This work will add more information in \cite{teepja}, where high spin states of $^{124-131}$Te isotopes are populated and 
shell model results on $^{128-131}$Te isotopes reported.

The broad conclusions from the present work are following:
\begin{itemize}
 
\item The yrast states of the four  $^{124-127}$Te isotopes are very well described by shell model.

\item From the  components of the wave functions, particularly the two values of $I_n$ and $I_p$ it is possible to identify
which nucleon pairs are broken to obtain the total angular momentum of the calculated  particular high-spin states.

\item The electric transition probabilities are in reasonable agreement with the experimental data with $e_{\pi} =1.5e$
and $e_{\nu} =1.0e $. 

\item  Results with the modification in $n-n$ part of the effective interaction show better agreement with the experimental data for low-lying states.

\item High-spin states in nuclei $Z$ $\sim$ 50 are expected from breaking of neutron/proton pairs. Experimentally it is
difficult to detect the de-excitation of these high-spin states through long-lived isomers. For theory
it is an ideal test of the two-body matrix elements of the residual interactions to reproduce these states.

\end{itemize}  

{\bf{Acknowledgments}}\\
One of the authors (P.C.S.) would like to thank to Prof. B. A. Brown for useful discussions.
VK, acknowledge financial support from CSIR, India for his thesis work. MJE acknowledges support from TWAS-CNPq grant and grant
F2-FA-F177 of Uzbekistan Academy of Sciences. Authors are grateful to Prof. Jesus Lubian for his help to prepare the manuscript.



\begin{thebibliography}{99}

\bibitem{nature}
K.L. Jones, {\it et~al.,} \emph{ Nature (London)}  \textbf{465}, 454 (2010). 


\bibitem{prc1}
 O. Sorlin and M.-G Porquet, \emph{ Prog. Part. Nucl. Phys.}  \textbf{61}, 602 (2008). 
 
 \bibitem{prc2}
 L. Coraggio, A. Covello, A. Gargano, N. Itaco, T.T.S. Kuo  \emph{Phys. Rev. C} \textbf{91}, 041301(R) (2015). 
 
\bibitem{prc3}
 L. Coraggio, A. Covello, A. Gargano, N. Itaco  \emph{Phys. Rev. C} \textbf{87}, 034309 (2013). 
 
\bibitem{prc4} 

H. -Kui Wang, Y. Sun, H. Jin, K. Kaneko, S. Tazaki, \emph{Phys. Rev. C} \textbf{88}, 054310 (2013). 

\bibitem{wood}
J.L. Wood, K. Heyde, W. Nazarewicz, M. Huyse, and P. Van Duppen
  \emph{Phys. Rep.} \textbf{215}, 101 (1992). 
 
\bibitem{prl1}
J. Taprogge, {\it et~al.,}  \emph{Phys. Rev. Lett.} \textbf{112}, 132501 (2014).

\bibitem{prl2}
G. S. Simpson, {\it et~al.,}  \emph{Phys. Rev. Lett.} \textbf{113}, 132502 (2014).

\bibitem{prl3}
H. Watanabe, {\it et~al.,}  \emph{Phys. Rev. Lett.} \textbf{113}, 042502 (2014). 





  






  

 
\bibitem{talmi}
I. Talmi, Simple Models of Complex Nuclei (Harwood Academic, Chur, Switzerland, 1993).


\bibitem{astierconf}
A. Astier,  \emph{J. Phys.: Conf. Ser.} \textbf{420},
  012055  (2014). 

\bibitem{teepja}
A. Astier, M.-G Porquet, Ts. Venkova  {\it et~al.,} \emph{The European Physical Journal A -
  Hadrons and Nuclei}\textbf{50}, 2  (2014).
  
 \bibitem{zhang}
C.T. Zhang et al.  \emph{Nucl. Phys. A}  \textbf{628}, 386 (1998). 

\bibitem{PhysRevC.85.064316}
A. Astier, M.~-G. Porquet, Ts. Venkova  {\it et~al.,}  \emph{Phys. Rev. C} \textbf{85},
  064316 (2012).

\bibitem{JPGN82}
P.~C. Srivastava,  M. J. Ermamatov, I. O. Morales, \emph{Journal of Physics G: Nuclear and
  Particle Physics} \textbf{40}, 035106  (2013).
  

\bibitem{PhysRevC.84.014329}
K. Wimmer, U. K\"oster, P. Hoff   {\it et~al.,}   \emph{Phys. Rev. C}  \textbf{84}, 014329 (2011).

\bibitem{PhysRevC.87.054316}
A. Astier, M. -G. Porquet,   {\it et~al.,}   \emph{Phys. Rev. C}  \textbf{87}, 054316  (2011). 
  
\bibitem{Morales2011606}
I.~O. Morales, P.~V. Isacker, I. Talmi, \emph{Phys. Lett.} \textbf{703B},
  606  (2011).
  
\bibitem{Mac}
R. Machleidt, F. Sammarruca, Y. Song, \emph{Phys. Rev. C} \textbf{53},
  R1483 (1996).
 
  
\bibitem{PhysRevC.71.044317}
B.~A. Brown, N.~J. Stone, J.~R. Stone, {\it et~al.,}  \emph{Phys. Rev. C} \textbf{71},
  044317 (2005).
  

\bibitem{MSU-NSCL}
B. A. Brown, W. D. M. Rae, E. McDonald,  M. Horoi, 
NuShellX@MSU.

\bibitem{nndc}
ENSDF dadabase, [http://www.nndc.bnl.gov/ensdf/].

\bibitem{broda}
R. Broda et al.,   \emph{The European Physical Journal A -
  Hadrons and Nuclei}  \textbf{20}, 145  (2004).
  
\bibitem{Pietri}
S. Pietri et~al.,  \emph{Phys. Rev. C} \textbf{83}, 044328  (2011). 

\bibitem{124}
T. Tamura,  \emph{Nucl. Data Sheets} \textbf{108}, 455  (2007). 


\bibitem{125}
J. Katakura,  \emph{Nucl. Data Sheets} \textbf{112}, 495  (2011). 

\bibitem{126}
J. Katakura, K. Kitao, \emph{Nucl. Data Sheets} \textbf{97}, 765  (2002). 

\bibitem{127}
A. Hashizume, \emph{Nucl. Data Sheets} \textbf{112}, 1647  (2011).


\bibitem{128}
M. Kanbe, K. Kitao, \emph{Nucl. Data Sheets} \textbf{94}, 227  (2001).

\bibitem{129}
J. Timar, Z. Elekes, \emph{Nucl. Data Sheets} \textbf{121}, 143  (2014).

\bibitem{130}
B. Singh, \emph{Nucl. Data Sheets} \textbf{93}, 33  (2001).

\bibitem{131}
Yu. Khazov, I. Mitropolsky, A.A. Rodionov, \emph{Nucl. Data Sheets} \textbf{107}, 2715  (2006).

\bibitem{132}
Yu. Khazov, A.A. Rodionov, S. Sakharov, \emph{Nucl. Data Sheets} \textbf{104}, 497  (2005).

\bibitem{133}
Yu. Khazov and A. Rodionov \emph{Nucl. Data Sheets} \textbf{112}, 855  (2011).

\bibitem{Lozeva}
R.L. Lozeva, G.S. Simpson, H. Grawe {\it et~al.,}  \emph{Phys. Rev. C} \textbf{77},
  064313 (2008).

\bibitem{ss2}
S. Sarkar, M. Saha Sarkar, \emph{Phys. Rev. C}  \textbf{78}, 024308 (2008).   

\bibitem{thesis}
A. Ekstr\"om, \emph{ "Effective charges in the nuclei in the vicinity of $^{100}$Sn"}, PhD thesis, Lund University, Sweden (2009).

\bibitem{110Sn}
J. Cederk\"all, A. Ekstr\"om, C. Fahlander, A. M. Hurst, M. Hjorth-Jensen, {\it et~al.,}  \emph{Phys. Rev. Lett.} \textbf{98}, 172501 (2007).

\bibitem{PRL2}
C. Vaman, C. Andreoiu, D. Bazin, A. Becerril, B. A. Brown, {\it et~al.,}  \emph{Phys. Rev. Lett.} \textbf{99}, 162501 (2007).

\bibitem{raman}
S. Raman, C.W. Nestor, and P. Tikkanen, \emph{At. Data Nucl. Data Tables}  \textbf{78}, 1 (2001).

\bibitem{back}
T. B\"ack {\it et~al.,}  \emph{Phys. Scr.}  \textbf{T150}, 014003 (2012).

\bibitem{saxena}
M. Saxena et~al.,  \emph{Phys. Rev. C}  \textbf{90}, 024316  (2014).

\bibitem{doncel}
M. Doncel et~al., \emph{Phys. Rev. C}  \textbf{91}, 061304(R)  (2015).  

\bibitem{quesne}
C. Quesne, \emph{Phys. Lett.}  \textbf{31B}, 7  (1970). 

\bibitem{polls}
A. Polls, H. Muther, A. Faessler, T.T.S. Kuo, and E. Osnes, \emph{Nucl. Phys.}  \textbf{401A}, 124 (1983).  

\bibitem{zele}
V. G. Zelevinsky, \emph{Phys. At.  Nucl.}  \textbf{72}, 1107 (2009).  

\bibitem{poves}

A. Poves and A. Zuker, \emph{Phys. Rep.}  \textbf{70}, 235 (1981).

\bibitem{piet}
P. Van Isacker and I. Talmi, \emph{Eur. Phys. Lett.}  \textbf{90}, 32001 (2010).  
\end{thebibliography}
\end{document}